%% file: main_arxiv.tex
\documentclass{article}

\usepackage{arxiv}
\usepackage{bigfoot}
\usepackage{etoolbox}

\usepackage[utf8]{inputenc}
\usepackage[sorting=none,giveninits=true]{biblatex}
\usepackage{graphicx}
\usepackage{subcaption}
\usepackage{layout}
\usepackage{xcolor}
\usepackage{siunitx}
\usepackage{comment}
\usepackage{authblk}
\usepackage{amsmath,amsfonts,amssymb}
\usepackage[thinc]{esdiff}
\usepackage{hyperref}
\hypersetup{
    colorlinks=true,
    linkcolor=blue,
    filecolor=magenta,      
    urlcolor=cyan,
    pdftitle={High-Fidelity Numerical Modeling},
    pdfpagemode=FullScreen,
    pdfauthor={David S.~Hippocampus, Elias D.~Striatum}, 
    pdfkeywords={First keyword, Second keyword, More}
    }

\DeclareNewFootnote{AAffil}[arabic]
\DeclareNewFootnote{ANote}[fnsymbol]

\makeatletter
\patchcmd\maketitle{\def\@makefnmark{\rlap{\@textsuperscript{\normalfont\@thefnmark}}}}{}{}{}
\makeatother

\makeatletter
\def\thanksAAffil#1{
  \footnotemarkAAffil\protected@xdef\@thanks{\@thanks%
        \protect\footnotetextAAffil[\the \c@footnoteAAffil]{#1}}%
}
\def\thanksANote#1{%
  \footnotemarkANote%
  \protected@xdef\@thanks{\@thanks%
        \protect\footnotetextANote[\the \c@footnoteANote]{#1}}%
}
\makeatother

\renewcommand{\shorttitle}{}

\newcommand{\bbeta}{\boldsymbol{\beta}}
\newcommand{\bb}{\boldsymbol{b}}
\newcommand{\bB}{\boldsymbol{B}}
\newcommand{\bx}{\boldsymbol{x}}
\newcommand{\bX}{\boldsymbol{X}}
\newcommand{\by}{\boldsymbol{y}}
\newcommand{\bY}{\boldsymbol{Y}}
\newcommand{\bu}{\boldsymbol{u}}
\newcommand{\bOmega}{\boldsymbol{\Omega}}

\newcommand{\bK}{\boldsymbol{K}}

\newcommand{\bE}{\boldsymbol{E}}
\newcommand{\bmu}{\boldsymbol{\mu}}
\newcommand{\bA}{\boldsymbol{\mathcal{A}}}
\newcommand{\bG}{\boldsymbol{\mathcal{G}}}
\newcommand{\bSigma}{\boldsymbol{\Sigma}}

\newbox{\orcid}\sbox{\orcid}{\includegraphics[scale=0.06]{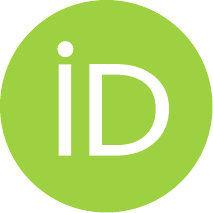}} 
\author[1]{%
	\href{https://orcid.org/0000-0001-8435-6466}{\usebox{\orcid}\hspace{1mm}Jacopo~Bonari\thanks{\texttt{jacopo.bonari@dlr.de}}}%
}
\author[1]{%
	\href{https://orcid.org/0000-0002-2814-0027}{\usebox{\orcid}\hspace{1mm}Max~von~Danwitz}
}
\author[1,2]{%
	\href{https://orcid.org/0000-0002-8820-466X}{\usebox{\orcid}\hspace{1mm}Alexander~Popp}
}

\affil[1]{German Aerospace Center (DLR), Institute for the Protection of Terrestrial Infrastructures, \newline 53757 Sankt Augustin, Germany}
\affil[2]{University of the Bundeswehr Munich, Institute for Mathematics and Computer-Based Simulation (IMCS), \newline 85577 Neubiberg, Germany}

\AtEveryBibitem{\clearfield{month}}
\AtEveryBibitem{\clearfield{day}}
\graphicspath{{figures/}}

\title{High-Fidelity Numerical Modeling for the Mechanical Characterization of a Full-Scale Test Bridge}

\begin{document}


\maketitle

\begin{abstract}
\input{abstract}
\end{abstract}

\section{Introduction}
Due to their contribution to economic activities, response to emergencies, and regional connectivity,
bridges represent critical components of transportation networks. Nevertheless, their functionality is increasingly threatened by a convergence of challenges that include rising traffic volumes and loads and the escalating frequency and intensity of extreme weather events. The former accelerates fatigue and deterioration processes happening on already aging structures, the latter is driven by climate change and exacerbates the aforementioned hazards. More specifically, intensified flooding events enhance riverbank and riverbed erosion, therefore posing severe threats to bridge foundations. Among these risks, scour, i.e., the localized erosion of soil around bridges piers and abutments due to flowing water, has emerged worldwide as one of the leading causes of bridge failures~\cite{zhang:2022}. If both natural factors, e.g., floods, and design errors are taken into account, then scour leads, in most of the cases, to catastrophic collapse, rather than partial failure~\cite{lee:2013}. Therefore, the implementation of robust SHM systems has become imperative to ensure the safety of bridge structures by means of an early detection of the damage, particularly in the context of increasing environmental and operational stresses.


Within this area of investigation, the analysis of the substructure response has received many attentions from researchers because of its conceptual and physical proximity to the problem~\cite{prendergast:2014}. On the other hand, the analysis of the superstructure might offer a robust alternative that avoids direct sensor placement in hazardous or out of reach zones and guarantee unhindered operational states even under severe environmental conditions~\cite{bao:2017}.

A major challenge in developing reliable SHM procedures is the scarcity of high-quality, real-world experimental data, particularly sets of data that include well-characterized damage states under realistic environmental and loading conditions. 
To tackle the problem and bridge this gap, a recent study from Jaelani et al.~\cite{jaelani:2023} 
introduced a new benchmark set of data collected from a two-span steel-concrete composite test bridge located at the University of the Bundeswehr Munich, and characterized by the enforcement of predefined damage scenarios and long-term ambient monitoring under real environmental and operating conditions, including the sample of ambient temperature variations.


The current study leverages this set of data and presents a SHM framework that analyzes and quantifies the settlements of the bridge central supports under ambient conditions. It combines the already mentioned high-quality experimental data with a detailed finite element (FE) model to perform Bayesian updating and refine the characterization of the structure, with a specific focus on the settlements of the central supports. The evaluation of key structural parameters and the quantification of their uncertainty is performed by incorporating the experimental observations into the FE model through probabilistic inference.

The results demonstrate the importance of high-fidelity structural models in enhancing condition assessment, enabling damage detection, and supporting the evolution of adaptive, intelligent monitoring systems. The unidirectional data stream from the physical structure to the numerical model constitutes a key step toward realizing a digital twin (DT), where the model serves as a digital shadow~\cite{brucherseifer:2021}, i.e., a dynamic, data-driven representation that evolves in response to incoming measurements and lays the groundwork for future bidirectional interaction~\cite{danwitz:2023,torzoni:2024}.


The reminder of the article is structured as follows. Section~\ref{sec:test_bridge} describes the test bridge present at the University of the Bundeswehr Munich and the campaign for experimental data acquisition; Section~\ref{sec:bayesian} features a linear Bayesian updating of the model to quantify the uncertainty related to the state of the foundations and describes the related results; finally, Section~\ref{sec:conclusions} describes possible further improvements for the SHM framework addressed and draws the related conclusions.

\section{Full-scale Test Bridge}\label{sec:test_bridge}

\subsection{Campaign of Experimental Data Acquisition}

\begin{figure}
    \centering
    \begin{subfigure}{\textwidth}
       \centering
        \includegraphics[width=0.9\textwidth]{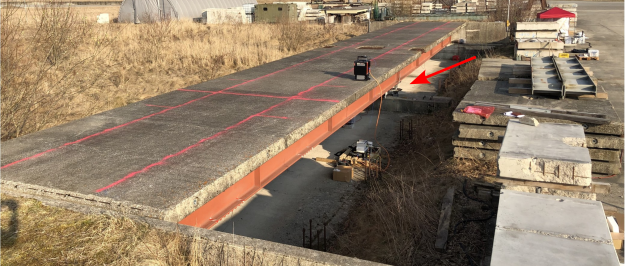}
        \subcaption{}
        \label{subfig:bridge_view}
    \end{subfigure}\\
    \vspace{5mm}
    \begin{subfigure}{\textwidth}
        \centering
        \includegraphics[width=\textwidth]{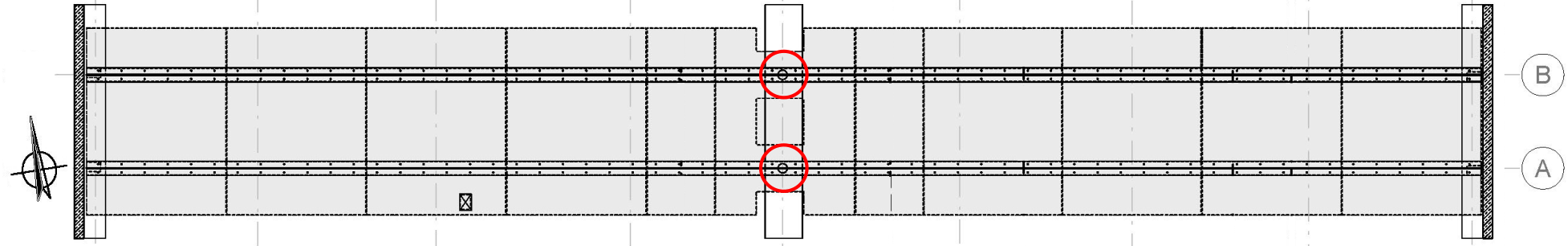}
        \subcaption{}
        \label{subfig:plan_view}
    \end{subfigure}
    \caption{Global view of the structure, the concrete segments that make the top deck are clearly visible, together with a portion of the southern main steel girder~(\subref{subfig:bridge_view}). The \emph{E-W} orientation of the test bridge can be inferred from its plan view~(\subref{subfig:plan_view}). The two central temporary supports are both highlighted in~(\subref{subfig:bridge_view}), by the red arrow, and in~(\subref{subfig:plan_view}), by red circles, with the southern support on the bottom side of the plan view and the northern on the top side. The pictures are displayed under Authors' permit from Jaelani et al.~\cite{jaelani:2023}.}
    \label{fig:test_bridge}
\end{figure}

The test bridge, located at the University of the Bundeswehr Munich, is a $30\,\si{\meter}$ long steel-concrete composite structure with two main girders and a segmented concrete deck, Fig.~\ref{fig:test_bridge}. In previous studies, it was used to investigate how several off-the-shelf smartphones and their built-in accelerometers could be used to assess the structural integrity of the construction~\cite{benndorf:2016,benndorf:2017}. The bridge features a flexible central support system that allows for controlled simulation of foundation settlements and several other different design damage scenarios. For a complete overview of all of them, the interested reader is redirected to Jaelani et al.~\cite{jaelani:2023}, while the focus of the present work is on the analysis and characterization of the structure under ambient conditions, obtained during a campaign of data acquisition that spanned, overall, almost one month.


\begin{figure}
    \centering
    \includegraphics[width=\textwidth]{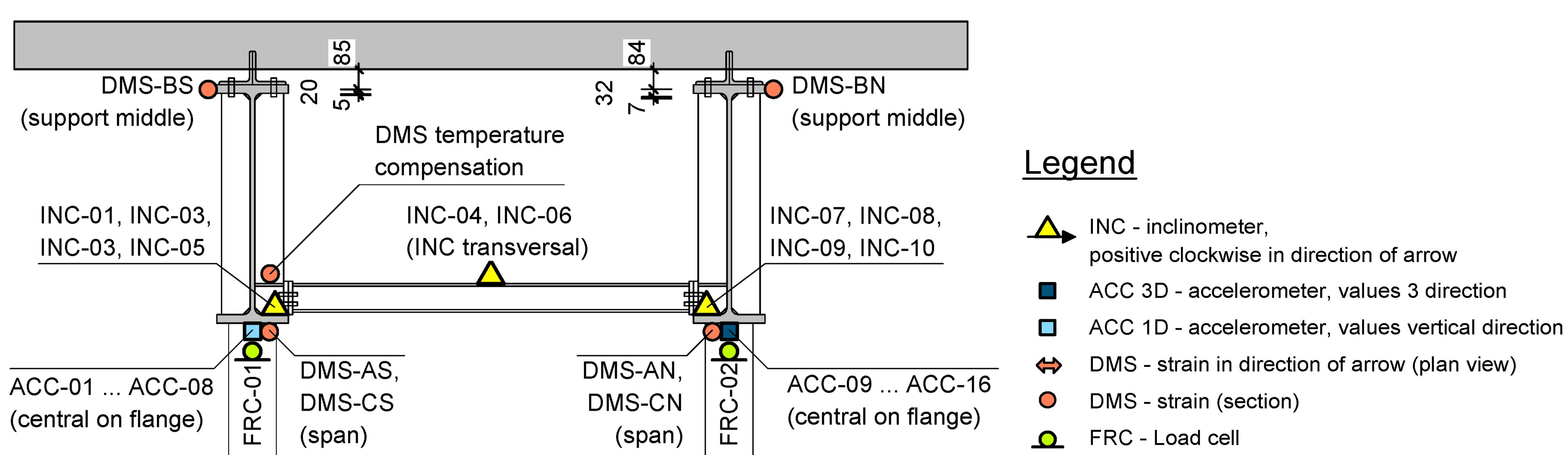}
    \caption{Sensors layout. The contribution focuses on results coming from the two load cells~\emph{FRC-01} and~\emph{FRC-02}, placed under the southern and northern side of the central support, and delivering the correspondent values $F_1\,(\si{\kilo\newton})$ and $F_2\,(\si{\kilo\newton})$ of the reaction forces of the bearings. Out of completeness, all the other sensors present in the same section are shown in the blueprint, while their type is explicated in the Legend.}
    \label{fig:sensors_layout}
\end{figure}

In this period, a mobile measurement system was deployed and data were collected via both short-term dynamic and long-term ambient monitoring, the latter spanning a time window of almost three weeks. An excerpt of the sensors layout employed, with reference to the central section of the bridge, can be observed in Fig.~\ref{fig:sensors_layout}. Throughout this contribution, the attention will be focused on sensor readings of reaction forces coming from the load cells directly placed under the two central supports and resulting in the quantities $F_1~(\si{\kilo\newton})$ and $F_2~(\si{\kilo\newton})$, for the southern support and the northern support, respectively. The analysis of long-term data reveals significant temperature-dependent variations, confirming that environmental effects must be taken into account in the analysis. More specifically, it can be observed how the southern support carries a higher load. Since the bridge lies in a favourable \emph{E-W} direction, cf. Fig.~\ref{subfig:plan_view}, and no surrounding elements interfere with solar exposure, 
it can be concluded that the southern part of the structure only receives direct sunlight, while the northern always lies in the shadow. In the current study, only air temperature values sampled in correspondence of the structure have been taken into account, while, on the other hand, a quantitative analysis of the solar irradiance~\cite{iea:2024} would be a more critical factor in accounting for thermal-induced structural response, since it directly drives surface heating and therefore it is the primary source of temperature differences that lead to thermal deformation across the structure; this kind of analysis will be the subject of further investigations.

\begin{figure}
    \centering
    \includegraphics{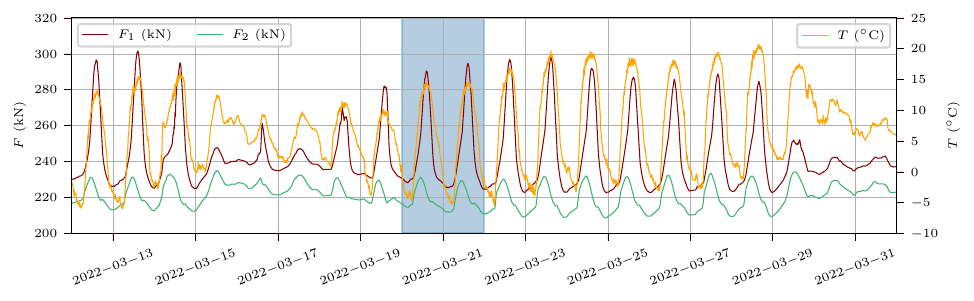}
    \caption{Reaction forces over the long term sampling. The shadowed region refers to a two-day period where a subset of the data will be extracted for further analyses. Data have originally been sampled at a frequency $f=10^3\,\si{\hertz}$ for the whole period. The current representation features a downsampling and consequent aggregation of the measurement over an interval $\Delta_T=60\,\si{\second}$.}
    \label{fig:time_temp_force}
\end{figure}

\subsection{High-Fidelity Finite Element Model}
As a complement of the real-structure, a high-fidelity FE model has been developed to enhance the value and utility of the data set, serving as a digital shadow~\cite{brucherseifer:2021} of the physical asset. By simulating the response behavior under the same conditions of the real structure, i.e., undamaged and engineered damage states, the virtual results can be compared to the actual experimental data and used in two different directions, i.e., both to better explain the observations and improve the characterization of the model. This allows, for example, a thorough examination of the redistribution of the reaction forces and the subsequent modified deformation status and stress condition resulting from differential foundations settlement.

During the modeling stage, a balanced trade-off between accuracy and computational efficiency has been sought. This resulted in an intermediate level of accuracy achieved through the consistent use of shell elements for all the primary structural components, this being the two main HEB1000 steel girders, their stiffeners, the deck concrete slabs, the HEB280 and HEB120 transversal braces, and the steel connectors 
linking the slab to the main girders, for which halved HEA340 profiles with incomplete circular holes were employed. To model the interaction between the segmented concrete slabs that compose the deck, a system of distributed springs was implemented. 
A critical and to some extent questionable modeling aspect lies in the representation of the bolted connections between the HEB1000 girders and the HEA340 profiles drown in the concrete. While the beam elements used to model these connections are computationally efficient and suited for modeling linear shear transfer, they represent a substantial simplification of the actual complex three-dimensional behavior of a bolted joint, that involves high local stress concentrations, microslips, friction, and prestress. Nevertheless, the choice reflects a pragmatic compromise ensuring that the model captures the essential composite action and load transfer, while keeping the overall model size and solution time manageable in the context of parametric studies and SHM analysis workflows. 

\begin{figure}
    \centering
    \begin{subfigure}{0.45\textwidth}
        \centering
        \includegraphics[width=0.8\textwidth]{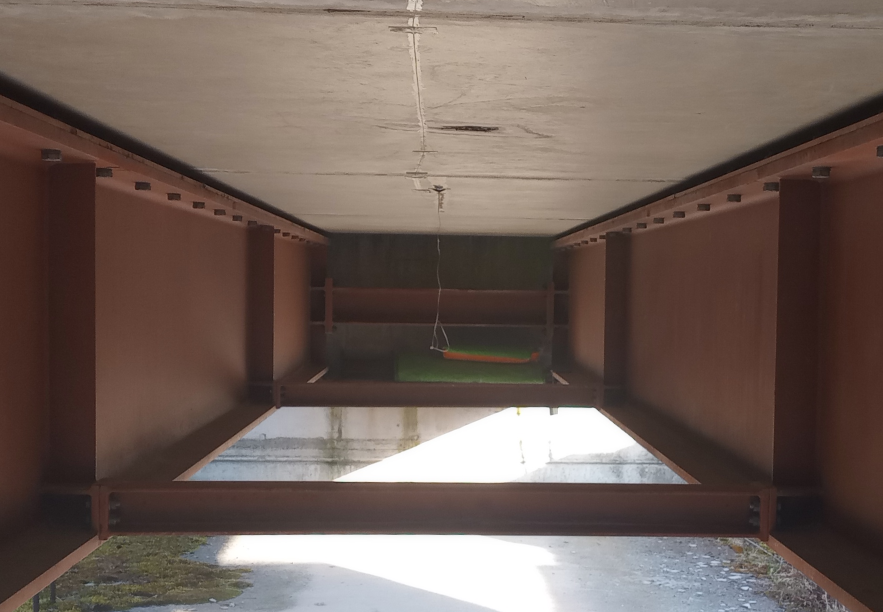}
        \subcaption{}
        \label{subfig:view_under_real}
    \end{subfigure}
    \hfill
    \begin{subfigure}{0.45\textwidth}
        \centering
        \includegraphics[width=0.8\textwidth]{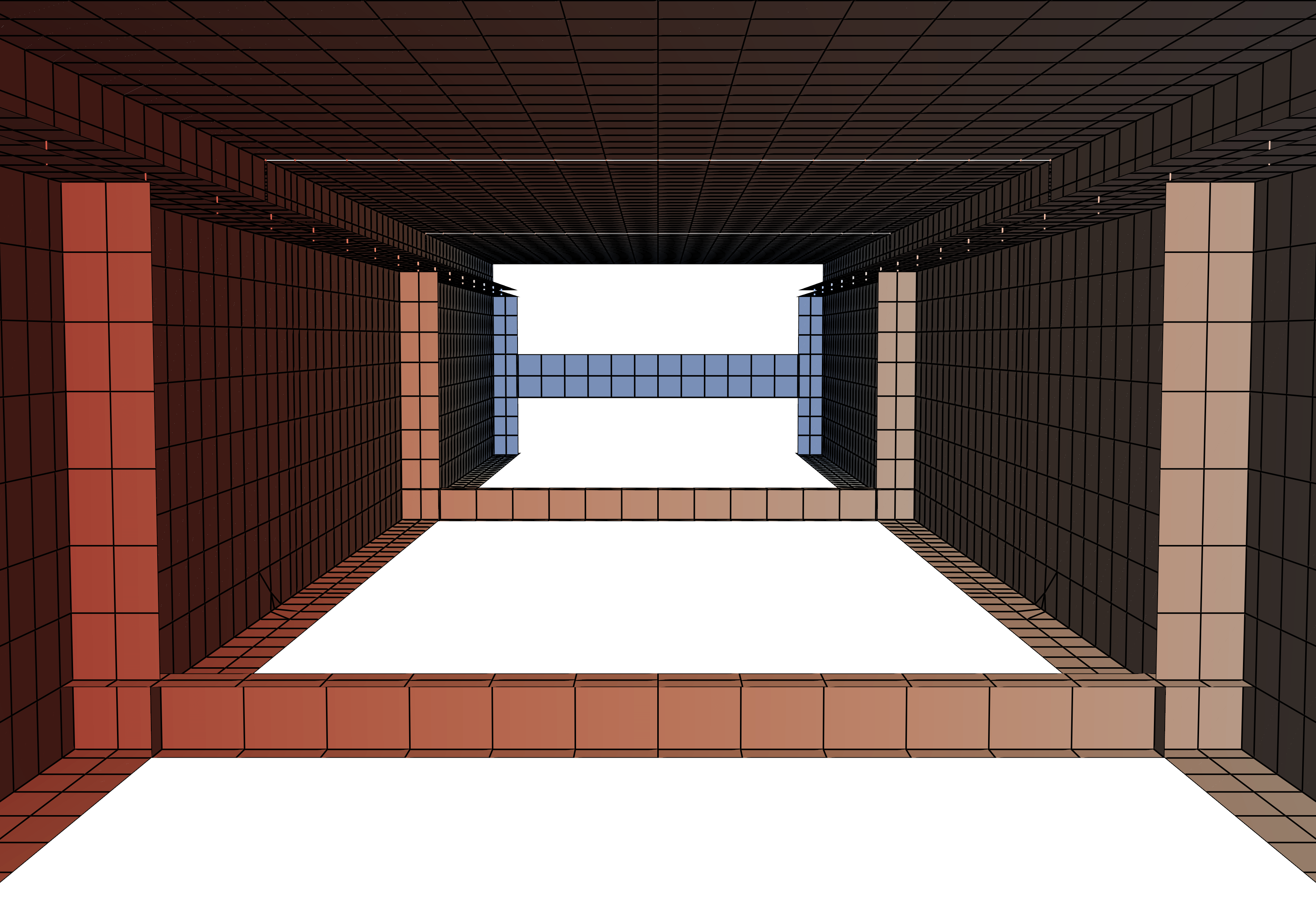}
    \subcaption{}
        \label{subfig:view_under_model}
    \end{subfigure}
    \\
    \begin{subfigure}{0.45\textwidth}
        \centering
        \includegraphics[width=0.8\textwidth]{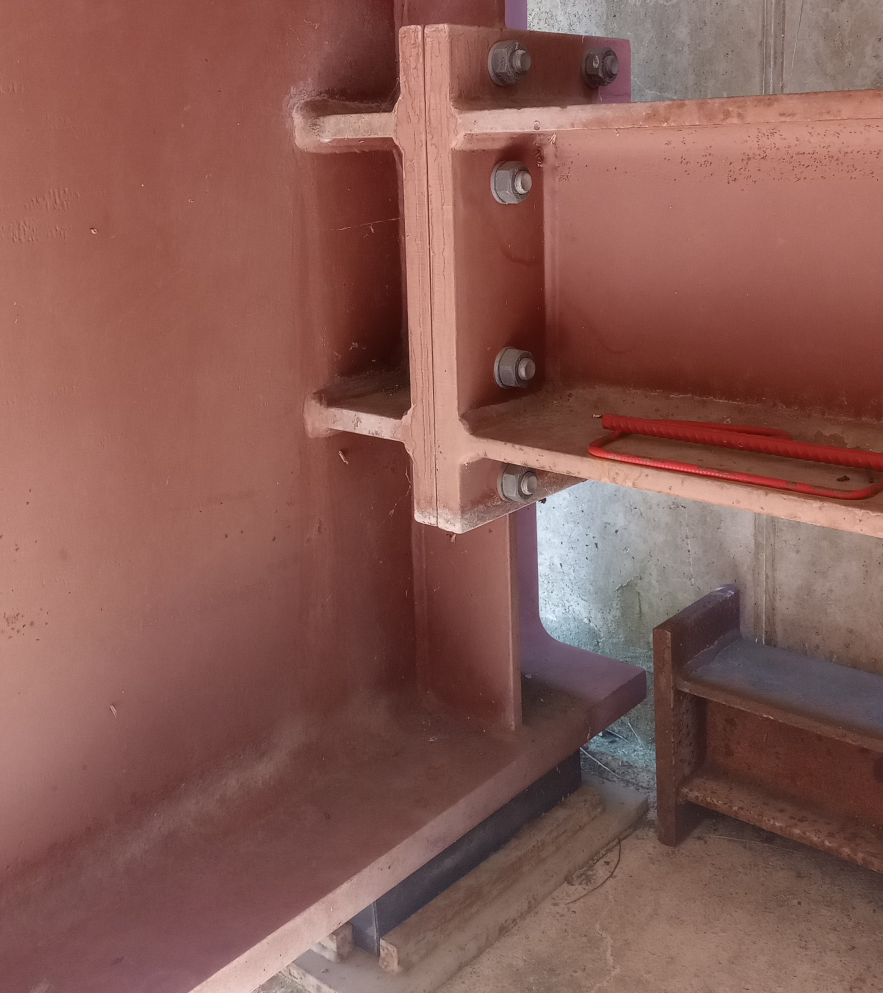}
        \subcaption{}
        \label{subfig:detail_real}
    \end{subfigure}
    \hfill
    \begin{subfigure}{0.45\textwidth}
        \centering
        \includegraphics[width=0.8\textwidth]{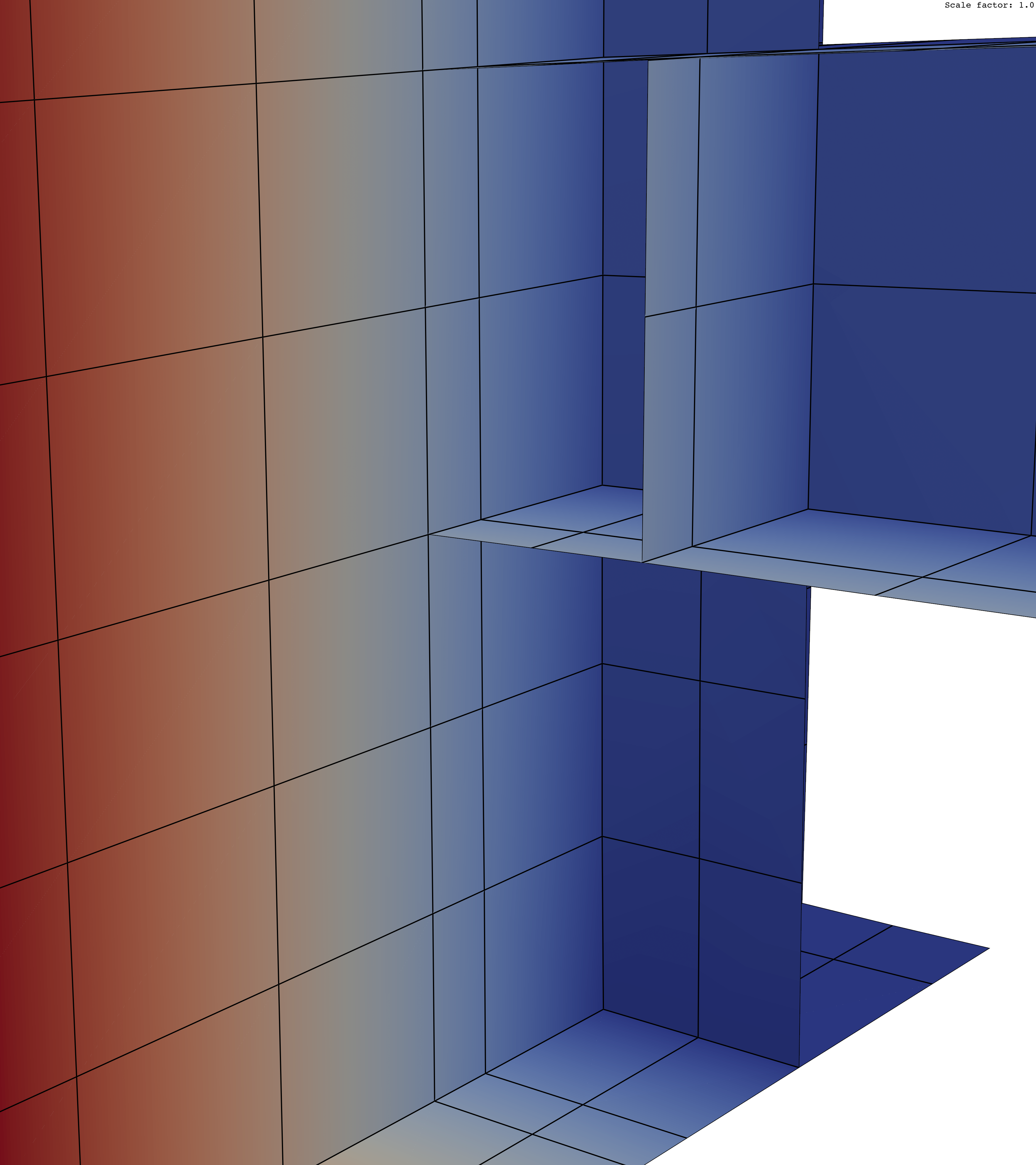}
        \subcaption{}
        \label{subfig:detail_model}
    \end{subfigure}
    \caption{Different views of the real structure set side by side to the FE model. Figure~(\subref{subfig:view_under_real}) shows a view of the structure as seen from below the deck, with bolts connecting the HEB1000 profiles to the concrete slabs in clear sight and highlighted by a red arrow; Figure~(\subref{subfig:detail_real}) presents a close look-up of the torsional stiffener set in correspondence of the eastern abutment, consistent in a flange coupling connecting the two main HEB1000 girders to a horizontal HEB280 profile. In parallel, Figs.~(\subref{subfig:view_under_model}) and~(\subref{subfig:detail_model}) represent the virtual replicas defined in the model.}
    \label{fig:struct_det}
\end{figure}

Based on an earlier FE model of the bridge, developed in ANSYS APDL and provided by Alexander Mendler~\cite{hansen:2023}, the FE model was  recreated in PyAnsys\footnote{Python interface to ANSYS; the full documentation is available at: \url{https://docs.pyansys.com/}.} to enable seamless integration in state-of-the-art data science workflows and a full automation of end-to-end pipelines starting with model setup down to results extraction, passing through parameter sweeps and simulation execution. For the specific case examined, the model has been linked to the experiments via a script that reads the measurement coming from the sensor, maps it to the FE model respective nodes, and performs direct comparison. Comparative views of the physical structure and the correspondent parts of the model can be observed in Fig.~\ref{fig:struct_det}.

\section{Model Calibration and Mechanical Characterization via Linear Bayesian Updating}\label{sec:bayesian}

The first step in processing the experimental data involved mitigating the influence of temperature on the measured reaction forces. 
In the case under examination, temperature significantly affects structural behavior inducing thermal deformation leading to spurious diverging variations in reaction forces, that mask the true signals. 
Observing the resulting data, it can be observed how a marked discrepancy is still present, whose source can not be a mere effect of daily temperature variations. Under the hypothesis that this difference is due to an undesired, parasitic differential settlement of the supports, the following section will leverage a surrogate model  of the linear relationship between forces and settlements to solve an inverse problem whose solution adds statistical knowledge about the foundation displacements that lead to the differential reactions at the supports, as per revealed by the data.

\subsection{Temperature compensation}
To mitigate the effect on the structure of daily temperature variations, a temperature compensation strategy was applied as presented in Neumann and Gertheiss~\cite{neumann:2022} and Neumann et al.~\cite{neumann:2025}, considering a temporal subset of the data acquired, cf. the shadowed area in Fig.~\ref{fig:time_temp_force}. It is assumed that $Y$ and $T$ are two random variables describing a sensor output and a potential confounder (in this case, reaction force at a single support and temperature, respectively), together with their realizations $y$ and $t$. If $y$ is regressed on $t$, then the signal covariance $\sigma^2_Y$ can be evaluated as the residual of the regression, an approach known in literature as partial covariance evaluation. If the non-linear regression function $y=g(t)$ is introduced, the sensor reading can be interpreted as:
\begin{equation*}
y = g(t)+\varepsilon,
\end{equation*}
where the regression function takes the form of the cubic B-spline~\cite{eilers:1996}:
\begin{equation}
g(t) = \sum_{k=1}^q b_k(t)\beta_k.
\label{eq:bspline}
\end{equation}

In Eq.~\eqref{eq:bspline} above, $[\beta_1,\dots,\beta_q]^\intercal=\bbeta$ are the basis coefficients that must be determined from the data, while $[b_1,\dots, b_q]^\intercal=\bb$ are pre-defined basis functions. Given a set of $m$ measured data points, the basis coefficients $\beta_k$ can then be identified through a least-squares minimization process. Since there is a temporal lag between the onset of temperature and the resulting response from the structure, the correlation results noisy and poorly defined, thus making the B-spline interpolation procedure unstable and prone to fit the asynchronous dynamics, resulting in an over-fitting that misses the global trend. The exploration of more complex regression algorithm tailored to tackle this specific problem, like dynamic time warping~\cite{mueller:2007} is left for further investigations. Here, to mitigate the issue, a smoothing term based on the integral of the square of the spline curvature is introduced as:
\begin{equation*}
\lambda \int_\tau g''(t)^2\,\mathrm{d}t = \lambda \bbeta^\intercal\bOmega\bbeta,
\end{equation*}
where $\bOmega$ is a basis dependent coefficient matrix and $\lambda$ a non-negative smoothing parameter. The solution of the penalized least squares problem can then be expressed in closed form as:

\begin{equation*}
    \hat{\bbeta} = \frac{\bB^\intercal\bx}{\bB^\intercal\bB+\lambda\bOmega},\quad\text{with}\quad\bB \in \mathbb{R}^{m\times q} : B_{i,k} = b_k(t_i).
\end{equation*}

\begin{figure}
    \centering
    \includegraphics[width=0.495\textwidth]{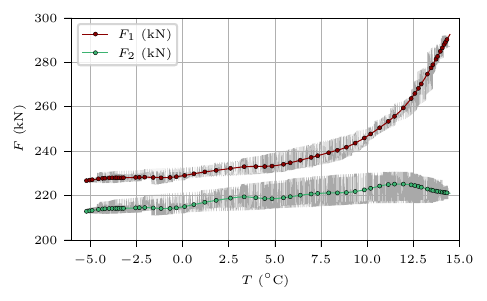}
    \caption{Functional relationship between temperature and reaction forces at the supports. The proposed penalized B-spline regression for $\hat{\lambda}=10$ is shown by circular markers for the southern, red curve, and northern, green curve support reactions. The perfect fit obtained for $\lambda=0$ is added for completeness (dense thin gray lines). It can be observed how the effect of temperature is marked for the southern support as compared to the opposite northern.}
    \label{fig:temp_regression}
\end{figure}

\begin{figure}
    \centering
    \begin{subfigure}{0.495\textwidth}
       \centering
        \includegraphics[width=\textwidth]{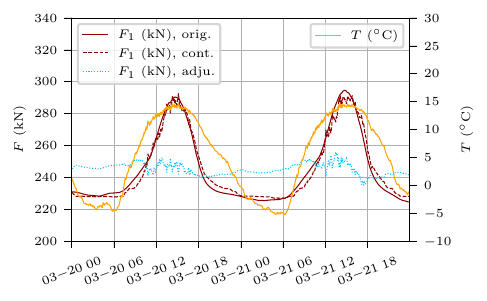}
        \subcaption{}
        \label{subfig:temp_detrend_a}
    \end{subfigure}
    \begin{subfigure}{0.495\textwidth}
        \centering
        \includegraphics[width=\textwidth]{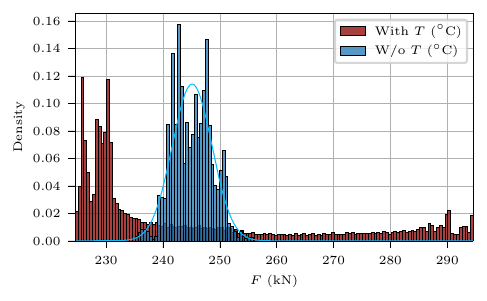}
        \subcaption{}
        \label{subfig:temp_detrend_b}
    \end{subfigure}
    \caption{Original reaction force at southern support, solid red line, together with the contribution to the load of air temperature oscillations, dashed red line. The adjusted force after the removal of the effect of the temperature is shown as a dotted blue line. Finally, temperature values are shown as a yellow solid line on the second vertical axis.~(\subref{subfig:temp_detrend_a}); probability density function of the reaction forces $F_1\,(\si{\kilo\newton})$ before and after the removal of the temperature effect~(\subref{subfig:temp_detrend_b}).}
    \label{fig:temp_detrend}
\end{figure}

The results of the regression with reference to the data presented in Fig.~\ref{fig:time_temp_force} can be seen in Fig.~\ref{fig:temp_regression} for two different values of the smoothing parameter $\lambda$. It can be observed how an (undesired) perfect fit can be obtained by imposing $\lambda=0$, represented as the gray shadowed dense thin lines in the background, while the final chosen $\hat{\lambda}=10$ result in a smooth fit that only captures the global trend. The regression function can now be subtracted from the raw data and added to the average value of the original signal, to isolate the structural response due to mechanical loading alone, Fig.~\ref{subfig:temp_detrend_a}. In an ideal case, i.e., for a perfect correlation between a unique confounder and the related output, the resulting signal should just represent the measurement error $\varepsilon$, and therefore be normally distributed. The different histograms for the distribution under examination, evaluated before and after the removal of the temperature effect can be observed in Fig.~\ref{subfig:temp_detrend_b}. This operation results in a more reliable and interpretable model.


\subsection{Surrogate model}
The finite element model has been systematically used to simulate the bridge response under various support conditions. Since the structural behavior is linear under the scenario considered, a surrogate model was derived to guarantee a fast mapping between foundation settlements and reaction forces. Specifically, the surrogate model takes the form
$\bu = \bA\bx +\bu_0$, where $\bu=[u_1,u_2]^\intercal$ are the reactions forces given by the numerical model, $\bx=[x_1,x_2]^\intercal$ the foundation settlements, and the superscripts 1 and 2 refer to the southern and northern central supports, respectively. Furthermore, $\bA$ is a generalized stiffness matrix and $\bu_0$ represents the reaction values in case of zero foundation settlement. To identify $\bA$ and $\bu_0$, the FE model has been solved for three distinct pairs of settlement values. In standard Bayesian inference notation, a relation between the surrogate model and the observation can be formalized introducing the observations/measurements $\by$ and the measurement error $\boldsymbol{\varepsilon}$:

\begin{equation*}
    \by = \bA\bx +\bu_0 + \boldsymbol{\varepsilon} = \boldsymbol{\mathcal{G}}(\bx)+\boldsymbol{\varepsilon}.
\end{equation*}



\subsection{Model calibration under ambient conditions}
Inferring foundation settlements from measured reaction forces is a classic linear inverse problem, and Bayesian model updating provides a principled framework to estimate the most probable settlement vector
$\bx^\star$ that best explains the data. The unknown settlement is now considered as a vector random variable $\bX$ with prior distribution $\pi_{\bX}(\bx) \sim \mathcal{N}(\mathbf{0},\bSigma_{\bX})$, with zero mean and a diagonal covariance matrix $\bSigma_{\bX} \in \mathbb{R}^{2\times2}$. The assumption that the prior has a zero mean is justified by the fact that, in the setting under consideration, settlements can also take positive values, i.e., the central supports can also be characterized by a positive mount; the independence of the two support settlements is rationalized in a diagonal covariance matrix. The equal covariance values on the diagonal have been chosen by applying the $3\sigma$ rule for the standard deviation of normal distributions and collocating the settlement value that makes the reaction force close to vanish on the threshold of the $99^\mathrm{th}$ percentile. With this assumption, the covariance matrix takes the values:
\begin{equation*}
    \bSigma_{\bX} = 
    \begin{bmatrix}
        (0.3\,\si{\milli\meter})^2 & 0.0 \\
        0.0 & (0.3\,\si{\milli\meter})^2
    \end{bmatrix}.
\end{equation*}
The Bayesian approach aims at updating the random vector $\bX$ given the observations $\by \in \mathbb{R}^{(m\times 2)}$ of the reactions, being $m$ the number of measurements. The posterior distribution $\bX'$ of the input random variable can be written as:
\begin{align}
    \pi_{\bX'}(\bx) &\propto \mathcal{L}(\by,\bx)\pi_{\bX}(\bx), &
    \mathcal{L}(\by,\bx) &= \prod_{i=1}^m \pi_E\bigl(\boldsymbol{\mathsf{y}}_i-\bG(\bx_i)\bigr),
    \label{eq:bayes}
\end{align}
where, employing the same notation that can be found in Bishop~\cite{bishop:2012}, a different typeface has been used to distinguish between a single observation of a multivariate output, $\boldsymbol{\mathsf{y}}_i \in \mathbb{R}^{(1\times 2)}$, and the array collecting the whole set of measurements, $\by$. In Eq.~\eqref{eq:bayes}, the error distribution $\pi_E = \mathcal{N}(\mathbf{0},\bSigma_{\bE})$ is assumed, again, to be characterized by zero mean and a covariance matrix $\bSigma_{\bE} \in \mathbb{R}^{2\times2}$ defined by the values:
\begin{equation*}
    \bSigma_{\bE} = 
    \begin{bmatrix}
        (6.04\times10^{-2}\,\si{\kilo\newton})^2 & 
        0.0 \\
        0.0 &
        (8.27\times10^{-2}\,\si{\kilo\newton})^2
    \end{bmatrix},
\end{equation*}
that have been directly inferred from the measurements. Given the linearity of the model, and since the parameters to be updated are Gaussian, the likelihood function can be expressed in closed form, and Bayes' rule re-cast as a quadratic minimization problem~\cite{rosic:2013}. In turn, this process allows to calibrate the statistical moments of the posterior distribution alone, instead of the whole probability distribution as would be necessary, for example, employing a Monte-Carlo method. The closed form expressions for the mean and covariance of the posterior distribution read, respectively~\cite{MARSILI2023110656}:

\begin{align*}
    \bmu_{\bX'} &= \bmu_{\bX} + \bK \bigl(\bar{\by}-\bG(\bmu_{\bX})\bigl), &
    \bSigma_{\bX'} &= \bSigma_{\bX}-\bK\bSigma_{(\bX,\bY)}^\intercal,
\end{align*}

where $\bSigma_{(\bX,\bY)}$ is the cross-covariance between the system inputs and the system outputs, and $\bK$ is the Kalman gain expressed by:

\begin{equation*}
    \bK = \frac{\bSigma_{(\bX,\bY)}}{\bSigma_{\bX}+\bSigma_{\bE}}.
\end{equation*}

\begin{figure}
    \centering
    \begin{subfigure}{0.495\textwidth}
       \centering
        \includegraphics[width=\textwidth]{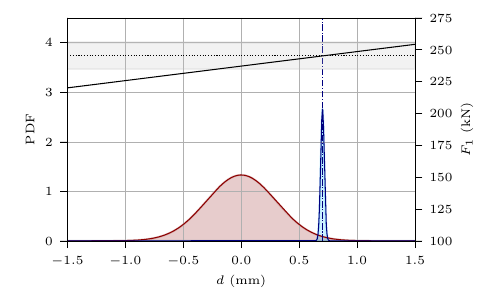}
        \subcaption{}
        \label{subfig:marginal_south}
    \end{subfigure}
    \begin{subfigure}{0.495\textwidth}
        \centering
        \includegraphics[width=\textwidth]{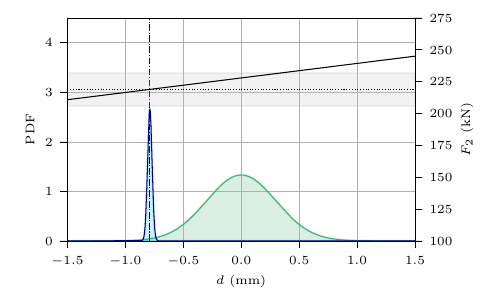}
        \subcaption{}
        \label{subfig:marginal_north}
    \end{subfigure}
    \\
    \begin{subfigure}{0.495\textwidth}
       \centering
        \includegraphics[width=\textwidth]{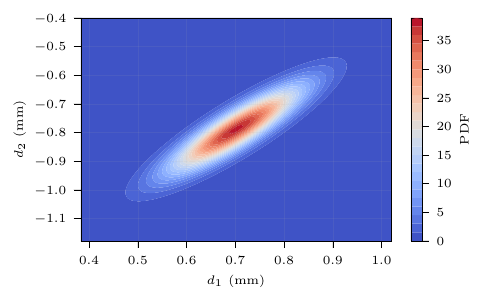}
        \subcaption{}
        \label{subfig:bivariate_posterior}
    \end{subfigure}

    \caption{Prior and posterior marginal probability density functions for support settlements on the southern~(\subref{subfig:marginal_south}) and northern side~(\subref{subfig:marginal_north}) of the structure. In both~(\subref{subfig:marginal_south}) and~(\subref{subfig:marginal_north}), it can be seen how the posterior mean intercepts the common locus of model response and experimental readings; here, the response from the model is shown as a solid black line, while data for $F_1$ and $F_2$ are shown by means of their average with a dotted line, and by the $\pm 3\sigma$ interval, with a gray shadowed area. Finally and out of completeness, the full bivariate PDF of the posterior distribution, now showing correlation between $d_1$ and $d_2$, is shown as well,~(\subref{subfig:bivariate_posterior}).}
    \label{fig:bayesian_update}
\end{figure}

The results of the calibration process are highlighted in Fig.~\ref{fig:bayesian_update}, where the posterior probability density function of the distribution of the the two settlement values is plotted against the prior. The results of the analysis hint that during the placement of the central support, a parasitic positive counter-mount of about $0.7\,\si{\milli\meter}$ was assigned to the southern support, while a negative undesired settlement of a similar magnitude was present in correspondence of the northern temporary bearing. Before the information provided by the data, the diagonal covariance $\bSigma_{\bX}$ reflected the completed independence of the inputs. The observation creates a statistical link between the two parameters, modeled by the introduction of off-diagonal terms in the posterior covariance $\bSigma_{\bX'}$ and resulting in bivariate normal distribution not aligned with the axes of the input parameter space, cf. Fig.~\ref{subfig:bivariate_posterior}.

\begin{figure}
    \centering
    \begin{subfigure}{0.495\textwidth}
       \centering
        \includegraphics[width=\textwidth]{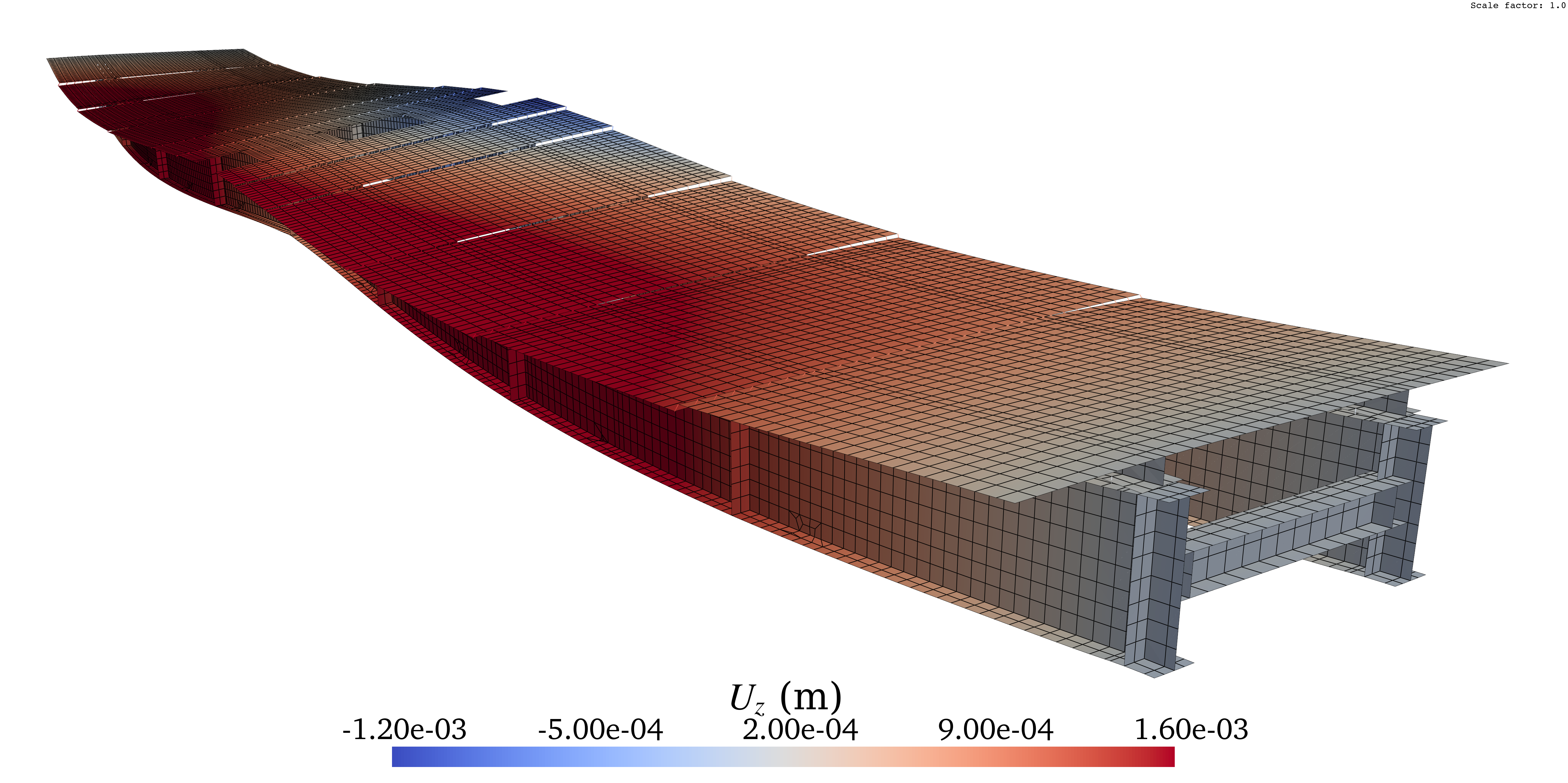}
        \subcaption{}
        \label{subfig:model_disp}
    \end{subfigure}
    \begin{subfigure}{0.495\textwidth}
        \centering
        \includegraphics[width=\textwidth]{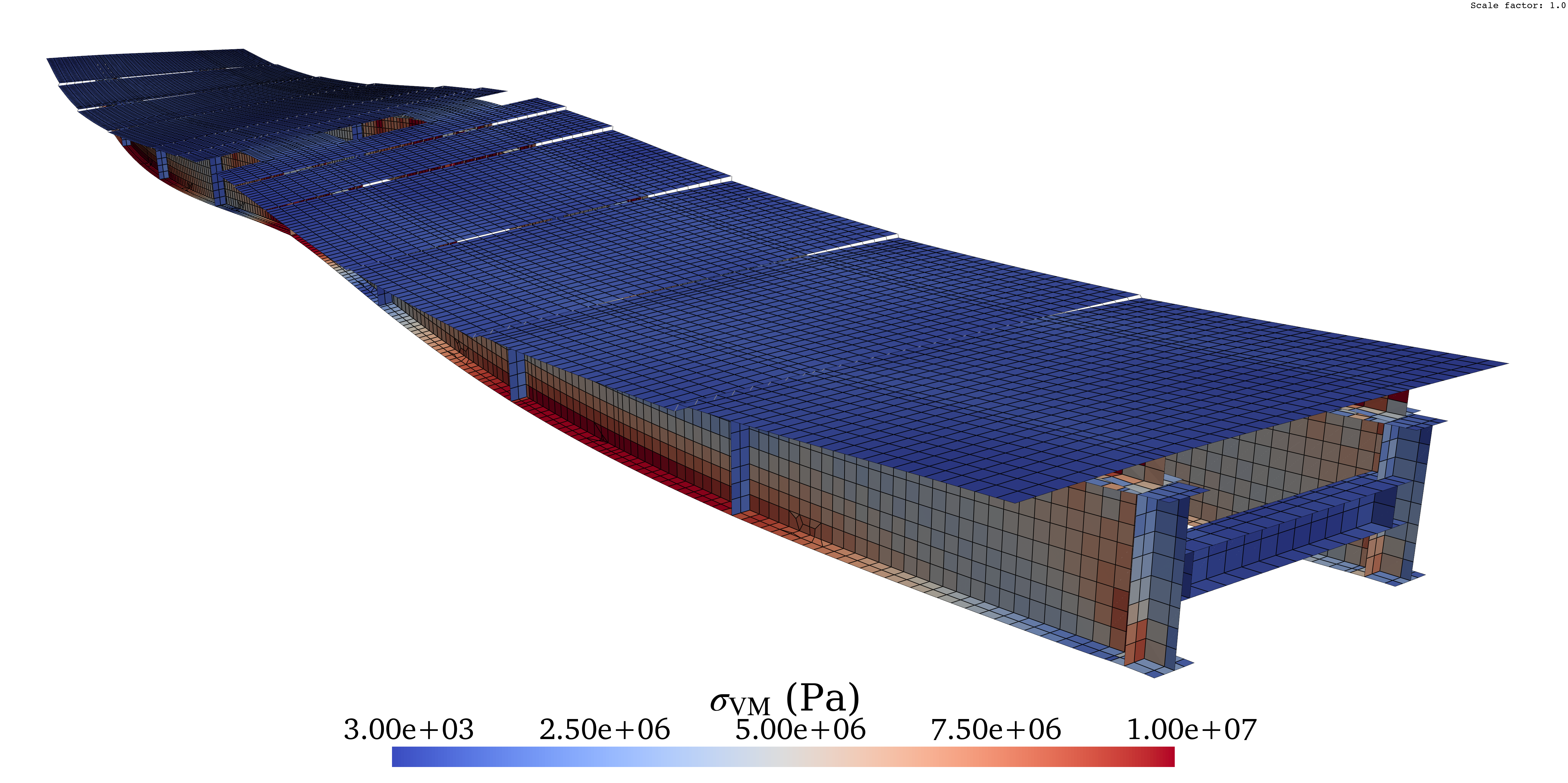}
        \subcaption{}
        \label{subfig:model_stress}
    \end{subfigure}
    \caption{Results of the FE model under the uneven support conditions identified by the solution of the inverse linear Bayesian problem. The solution is shown in terms of displacements,~(\subref{subfig:model_disp}) and von Mises stress,~(\subref{subfig:model_stress}). Both graphical representations have been scaled by a factor proportional to the displacement field to highlight the deformations.}
    \label{fig:final_model}
\end{figure}

Finally, a qualitative view of the structural response of the bridge after the characterization is shown in Fig.~\ref{fig:final_model}, with reference to the updated support conditions.



\section{Conclusions and Outlook}\label{sec:conclusions}
This study establishes a foundational framework for structural health monitoring (SHM) by integrating high-fidelity experimental data with a detailed finite element (FE) model, leveraging the benchmark dataset from Jaelani et al.~\cite{jaelani:2023} to investigate foundation settlements under ambient conditions. The approach combines three elements:~\emph{(i)} the use of a high-fidelity FE model to capture the structural behavior with physical accuracy;~\emph{(ii)} advanced computational techniques to account for environmental influences—particularly temperature-induced effects—on the measured response; and~\emph{(iii)} real-world data collected during a comprehensive, long-term monitoring campaign. The integration of these components significantly enhances both data quality and model reliability, enabling precise quantification of structural parameters through Bayesian model updating. The unidirectional data stream from the physical structure to the numerical model defines a digital shadow, i.e., a dynamic, data-driven representation that evolves in real time, forming the core of a DT concept. This synergy between physical and virtual assets not only improves the accuracy of condition assessment and uncertainty quantification but also paves the way for adaptive, intelligent monitoring systems capable of addressing the growing challenges posed by climate change and aging infrastructure. Ultimately, this integrated approach provides a robust, scalable foundation for advancing SHM frameworks in real-world bridge networks.


Further developments will proceed on the path laid down by this study. More advanced algorithms to subtract the environmental effects from the available data will be tested, after having identified dynamic time warping~\cite{mueller:2007} as suitable analysis method. For what concerns the available set of data, only a very small portion of the available information has been employed as of today. Future studies will expand their use, proceeding with the analysis of experimental information coming from a design scenario consisting in an artificial lowering of the central support by an amount way higher than the small parasitic foundation settlement found in the current study. This will set an additional challenge in case of nonlinear structural response, since more advanced model surrogation techniques, like generalized polynomial chaos expansion (gPCE)~\cite{sudret:2008} will have to be employed. Moreover, the presented Bayesian model updating procedure provides a principled framework to also include data of spaceborne monitoring capabilities~\cite{Scheiblauer.2025, Malinowska.2025} in the future.
Still keeping the focus on damages at the foundation level, also vibration based scour detection as described in~\cite{bao:2017} is considered a critical topic in the context of infrastructure protection with reference to natural hazards and worth of further investigations.

Beyond refinements and extensions of the proposed SHM pipeline for individual bridges, adopting a population-based perspective on similar structures could offer significant advantages in critical infrastructure protection~\cite{Gardner.2021}. Moreover, integrating structural and social vulnerability in bridge maintenance considerations enables a holistic infrastructure operation and resource prioritisation~\cite{MALINOWSKA2026106115}.

\printbibliography

\end{document}

%% file: abstract.tex
Bridges are vital components of transportation networks, serving as critical lifelines that ensure the safe and efficient movement of people, goods, and emergency services. With aging infrastructures, increasing traffic volumes and loads, and growing impact of extreme weather events driven by climate change, the development of reliable structural health monitoring (SHM) strategies has become of utmost importance. A key challenge in this domain is the scarcity of data on well-characterized damage states. To address this, a monitoring campaign was recently conducted on a full-scale, two-span test bridge specifically designed and built at the University of the Bundeswehr Munich to investigate damage scenarios related to specific structural deficiencies of the deck and to foundations settlement, the latter being connected to failure mechanisms typical in the context of floodings, when scour, i.e., washing out of the foundations, might happen. The test bridge is a crucial intermediate step between laboratory-scale experiments and real-world monitoring. In this study, a high-fidelity, physics-based numerical model of the same structure is presented as a complementary tool. The model enables accurate performances assessment and provides a detailed reference to interpret measured responses under varying environmental conditions and artificial damage scenarios. Experimental data collected under operational conditions were used to refine the model's mechanical characterization through Bayesian updating. The goal is to develop a functional digital twin of the test bridge, acting as a dynamic, data-driven shadow of the physical structure, to support informed maintenance decisions, extend service life, and enhance safety in future studies applied to real-world infrastructures.